\newcommand{\be}{\begin{eqnarray}}
\newcommand{\ee}{\end{eqnarray}}
\def\kms{km\thinspace s$^{-1}$\ }
\def\ms{m\thinspace s$^{-1}$\ }

\def\deg{$^\circ$}

\def\lgt {log~$\tau$}

\documentstyle[galley,epsf]{mn}
\begin{document}
\title[33 Lib ]{Radial Velocity Variations in Pulsating Ap Stars.\\
II. 33 Librae \footnote{Based on observations made at McDonald Observatory}}
\author[Mkrtichian et al.]{
D.E.  Mkrtichian$^{1,2}$, A.P. Hatzes$^3$,  \& A. Kanaan$^4$ \\
$^1$Astrophysical Research Center for the Structure and Evolution of the Cosmos, Sejong University, Seoul 143-747, Korea\\
$^2$Astronomical Observatory, Odessa National University, Shevchenko Park, Odessa, 65014, Ukraine\\
$^3$Th\"uringer Landessternwarte Tautenburg, Sternwarte 5, D-07778, Tautenburg,
Germany\\
$^4$Departamento de F\'{\i}sica, Universidade Federal de
Santa Caterina, Campus Universit\'ario, 88040-900, Florian\'opolis,
 Brazil \footnote {CNPq Fellow}
}
\date{}

\maketitle

\begin{abstract}

We present precise relative radial velocity (RV) measurements
for the rapidly oscillating Ap (roAp) star 33 Librae measured from high
resolution data spanning the wavelength interval 5000--6200 {\AA}. We
find that pulsational radial velocity amplitude determined over
a broad wavelength range ($\approx$ 100 {\AA})
depends on the spectral region that is examined
and can be as  high as 60 {\ms} at 5600 {\AA} and as low as 7 {\ms} in
the 5900 {\AA} region. RV measurements of individual spectral lines 
can show  higher amplitudes than results obtained using a 
``broad-band'' measurement that includes many spectral lines.
The acoustic cross-sections of the atmosphere, i.e. the phase and amplitude of 
the pulsations, as a function of optical depth 
is found for spectral lines of Ca, Cr, Fe, La, Ce, Gd, Er  and
Nd. This analysis shows that pulsation phase is variable through the
atmosphere and that Nd III lines pulsate almost 180$^\circ$ out-of-phase with
those of Nd II features and are
formed  significantly higher in the stellar atmosphere.
This conclusively establishes the presence of at least one radial
node to the pulsations in the upper stellar atmosphere.
We have estimated that this acoustic node  is located  above an optical
depth $log \, \tau < -$4.5 and below the level where the Nd III lines
are formed. We also suspect that there may be a second atmospheric node
in the lower atmosphere below,  or at {\lgt} $\simeq$ $-$0.9  and close
to continuum formation
level.

The histogram of pulsational phases 
for all individual spectral
feature shows a bi-modal Gaussian distribution with 17\% of the lines having a
pulsational phase $\approx$ 165{\deg} out-of-phase with most other
spectral lines. This is also consistent with the presence of a 
radial node in the stellar atmosphere.  The accumulation of phase
due to a running wave component can explain the 165{\deg} phase
difference as well as the broader width (by a factor of two) of one of the
Gaussian components of the phase distribution.

We also found evidence for phase variations as a function of effective
Lande g-factors. This may be  the influence of magnetic field and magnetic
intensification effect  on depths of spectral lines formation and that
magnetic field is controlling the pulsations. Our RV measurements  for
33 Lib suggest  that we are seeing evidence of vertical structure to the
oscillations as well as the influence of the distribution of elements on
the stellar surface.

We suggest and briefly discuss a  new semi-empirical tomographic procedure
for mono- and multi-mode roAp stars that will use acoustic cross-sections
obtained on different chemical elements and different pulsation modes for
restoring the abundance and acoustic profiles throughout
the stellar  atmosphere and across the stellar surface.

\end{abstract}

\begin{keywords}
Stars:individual:33 Lib -- Stars:pulsation -- Stars:variables 
\end{keywords}

\section{Introduction}
	The rapidly oscillating Ap stars (roAp) are a subclass of 
the  magnetic A stars that 
pulsate in high order, low-degree modes (see review
by Kurtz  1990). Much of our
knowledge about the pulsations in  roAp stars have come from photometric
studies since the short periods of the oscillations (5--15 min) make it
difficult to accumulate the  high quality spectral data needed for
radial velocity studies. Recently, radial  velocity
measurements have been made on several roAp stars and these are starting
to yield exciting results.

\begin{figure*}
\epsfxsize=14truecm
\epsffile{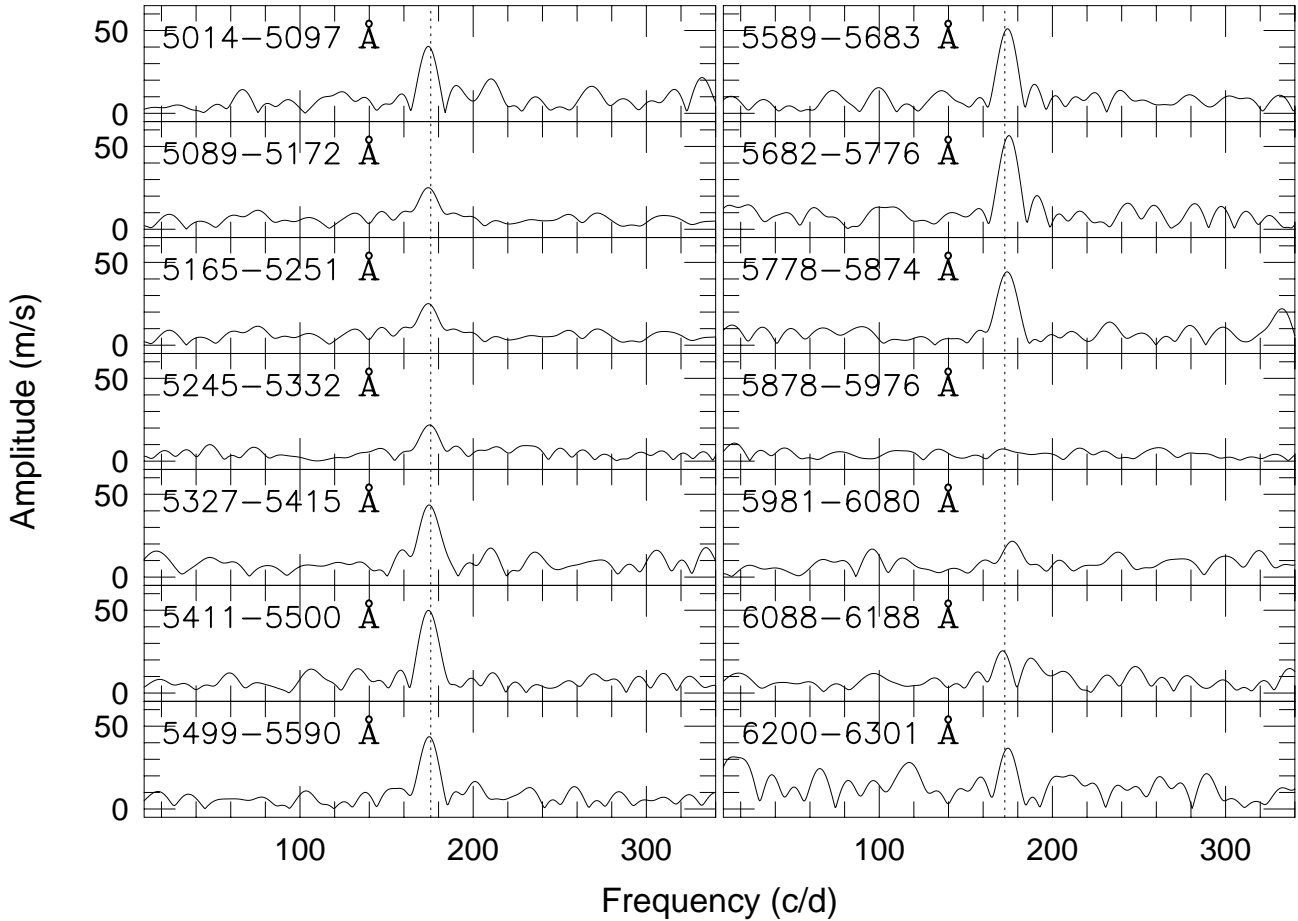}
\caption{The discrete Fourier transform of our RV measurements for 33 Lib.
Each panel represents a different spectral order with a wavelength range
specified by the numbers in the upper left of each panel.
The vertical dashed line is the
location of the principal photometric frequency.
\label{fig:dft}}
\end{figure*}

	Early investigations concentrated merely on detecting  radial velocity (RV)
variations in roAp stars primarily to confirm that pulsations were
indeed responsible for the photometric variations and to measure the RV
amplitude-to-photometric amplitude ratio (2$K$/$\Delta m$) which is
useful for comparing to values derived for other pulsating stars. The first
succesful attempt at measuring the pulsational RV for an roAp star
was done for HR 1217 by Matthews et al. (1988).
The estimated 2$K$-amplitude for this star over
the wavelength interval 4350 -- 4500 {\AA} was about
400 {\ms} which yielded a 2$K$/$\Delta m$ of 59$\pm$12 {\kms}\thinspace mag$^{-1}$,
a value comparable to that found in Cepheid-type variable
stars. There was evidence that the RV amplitude varied with rotational
phase since observations made on the previous  night showed no
significant RV variations above an amplitude of 130 {\ms}. Belmonte et
al. (1989) detected RV oscillations in HR 1217 with a frequency of 2.72
mHz and an amplitude of 258  $\pm$ 60 {\ms} This yielded a
2$K$/$\Delta$$m$ of $\approx$ 190 {\kms}\thinspace mag$^{-1}$, significantly
higher than the Matthews et al. (1988) result.

RV measurements at much higher RV precision  using a self-calibrating iodine
absorption cell were subsequently made for two other roAp stars.
Libbrecht (1988) measured an RV
amplitude of  about 42 {\ms} in $\gamma$ Equ based on observations
spanning the wavelength interval 5322 -- 5377\,{\AA}.  This resulted in
2$K$/$\Delta m$ = 26 {\kms}\thinspace mag$^{-1}$. Hatzes \& K\"urster
(1994) placed an upper limit of 23 {\ms} to any RV variations of the
roAp star $\alpha$ Cir based on spectral data spanning the wavelength
interval 5365 -- 5410 {\AA}. This yielded 2$K$/$\Delta m$ $<$ 10
{\kms}\thinspace mag$^{-1}$.

More detailed studies of roAp stars have established that the
pulsational RV behavior for these stars
can be quite complex with measured RV amplitudes differing by factors
of 10--100 for the same star depending on the spectral region that is
examined. 
In contrast to the low amplitude measured by
Hatzes \& K\"urster (1994),
Baldry et al. (1998) measured the RV
pulsational amplitude for $\alpha$ Cir and found that this 
was as high as   1 {\kms} in some spectral regions. Furthermore, those
spectral regions dominated by strong lines had a lower pulsational RV
amplitude than weaker spectral lines.

Kanaan \& Hatzes (1998; hereafter Paper I) presented RV
measurements for $\gamma$~Equ derived using a number of spectral  lines
which showed that the pulsational RV amplitude depended not only on line
strength, but on atomic species as well. Since weaker spectral lines are
formed, on average, deeper in the stellar atmosphere the amplitude
variations found in Paper I were interpreted as an atmospheric  height
effect resulting from a radial node situated in the stellar atmosphere.
The line-by-line RV measurements for $\gamma$~Equ  with more limited temporal
and wavelength coverage by 
Kochukov \& Ryabchikova (2001) confirmed the strong
dependence of pulsation amplitudes from atomic species and phase
differences for different elements and stages of ionization.

In retrospect these large variations
in the measured value of  2$K$/$\Delta$m in roAp stars are not
surprising. 
Mkrtichian (1992) first noted that the
spotty distribution of elements on roAp stars would greatly affect
the measurement of the RV amplitude and consequently the RV-to-photometric
amplitude ratio. A
2$K$/$\Delta$m determination for roAp stars would be meaningless since
it would depend on the spectral lines used for the RV determination.
This seems to be confirmed by recent RV studies of these stars.


Baldry et al. (1998)  found that approximately
15\% of the wavelength bands they examined in $\alpha$ Cir showed a 180$^\circ$ phase
shift to other bands. One explanation of such a bi-modal distribution 
is a horizontal node effect due to the inhomogeneous surface distribution
of elements (Mkrtichian 1992, 1994).  Alternatively, this bi-modal distribution
in phases can result from different lines pulsating in opposite phases
across a vertical atmospheric  radial node (Baldry et al. 1998). The
later is consistent with the amplitude variations in line strength found
by Kanaan \& Hatzes (1998) and the decline in photometric amplitude with
increasing wavelength (Medupe \& Kurtz 1998). Although these studies are
suggestive of the presence of such a wave, this is not conclusive.
Furthermore it may be difficult to disentangle the effects of the
vertical standing waves to those due to the horizontal surface
amplitude distributions of non-radial pulsations (NRP). Mkrtichian et al. (2000) suggested separating the
horizontal and vertical atmospheric effects by using spectral lines that
are formed at different atmospheric heights and that have a different
horizontal abundance distribution across the stellar surface. Understanding the
pulsations in roAp stars is indeed challenging, but the rewards may be
great in that we could derive important information about the
3-dimensional structure (vertical and horizontal)  of the pulsations.
RV measurements will play a key role in such studies.

We have started a program to use precise RV measurements to study the
pulsations in roAp stars. This program has already detected
``broad-band'' (i.e. computed over a large wavelength range) RV
variations in the roAp stars  HD~134214, HR~1217 (Hatzes et al. 1999a;
Hatzes et al. 2002),
33~Lib (Hatzes et al. 1999b),  and HD~122970 (Hatzes et al. 2000). A
line-by-line analysis using high resolution ($R$ (=$\lambda$/$\Delta
\lambda$)$>$ 45,000) data covering a wide wavelength range has been made
for relatively few roAp stars. If we are to understand the influence of
the magnetic field, surface distribution of elements, and atmospheric
structure on the characteristics of the pulsational RV variations then
we must make these kinds of measurements for more  of roAp stars
spanning the available parameter space covered by these stars. Here we
present a time series of precise RV measurements, both broad-band and
line-by-line, for the roAp star 33 Lib.

\begin{figure}
\epsfxsize=8.5truecm
\epsffile{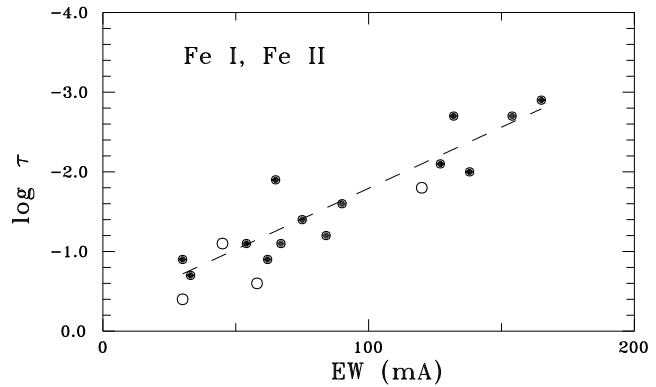}
\caption{The $W_{\lambda}$ -- {\lgt} transformation scale 
of the model atmosphere of 33~Lib for Fe I (dots) 
and Fe II (open circles).
\label{fig:few}}
\end{figure}
\section{The Star}

	33 Lib (=HD 137949) is a cool Ap star with an effective 
temperature of 7350 K (Ryabchikova et al. 1999) that has been classified
earlier as a Cr-Si-Eu star. An effective magnetic field strength
of $\approx$ $+$1600 Gauss has been measured for this star
which is only slightly variable (Van den Heuvel 1971; Wolff 1975).
Mathys et al. (1997) combined their data and all other available
magnetic field measurements for this star and found evidence for a
slow increase in the longitudinal field component at a rate of
$\sim$ 20 G yr$^{-1}$ which  implied a  rotational period $>$ 75 years.
	
	The roAp status for this star was established by Kurtz (1982)
who found a pulsational period of 8.272 min ($\nu_1$=2.01482 $mHz$). This star pulsates in
essentially one mode, although Kurtz (1991) found evidence for  a second
frequency at 1975 $\mu$Hz which has yet to be confirmed. That work also
detected the second harmonic at 2$\nu_1$ = 4.02956 $mHz$.

	Belmonte et al. (1989) attempted to search for RV variations
in 33 Lib using a Fabry-Perot interferometer and a stellar line
at 5317 {\AA}. Using more than 6 hours
of data they found  a peak in the power spectrum at the appropriate
frequency and with an amplitude of 310 {\ms}; however, due to the high
noise level in the power spectrum this result was uncertain.

\section{Observations}

	Observations of 33 Lib were made using the 2-d coud{\'e}
spectrograph (Tull et al. 1995) of the 2.7-m telescope at McDonald Observatory
on the night of 26 July 1997 between 3:15 UT and 5:26 UT.
One focus of the instrument provides a resolving power
$R$ (=$\lambda$/$\Delta \lambda$) = 60,000 with a nominal wavelength
coverage of 4000--10,000 {\AA} in one exposure when
used with a Tektronix 2048$\times$2048 CCD detector. Because of the
short pulsation period  of 33 Lib
it was necessary to minimize the time
between successive exposures. This was  accomplished by binning the CCD
by a factor of two perpendicular to the dispersion direction. This halved
the detector readout time without sacrificing 
spectral resolution. A further reduction in readout time was accomplished
by framing the CCD so as to record  only those pixels  on the detector
covering the spectral range 4700 -- 7000 {\AA}. This
binning and framing  of the CCD resulted
in a dead time of 15 secs (CCD readout and data storage). Exposure times
were 50 secs resulting in a full duty cycle of 65 secs. Observations
were made continuously one after another 
resulting in 86 independent observations.

\section{Results}
\subsection{Broad Band RV Variations}
	
	Radial velocities were first computed using each full
spectral order as these provided the best RV precision and resulted
in a fast determination of the mean RV amplitude over a broad spectral
region.
Figure~\ref{fig:dft} shows the Discrete Fourier
Transform (DFT) for the  RV measurements from each of the spectral orders.
The fundamental period, as marked by the vertical dotted line, is present in
every spectral order except the one spanning 5878 -- 6080 {\AA}.

The measured RV frequencies are all consistent, to within the
measurement errors, with the published photometric one. In the analysis
that follows we will assume that the 8.272 min period ($\nu$ = 2.0148
mHz = 174.08 c\,d$^{-1}$) is the only one that is present 
in all the spectral
orders. A least squares sine fit was made to the RV amplitude in each spectral
order keeping the period fixed at 8.272 min.

	Precise stellar radial velocity measurements were made using
an iodine absorption cell placed just before the entrance slit of the
spectrograph during each observation. 
This technique is commonly used 
for measuring precise relative stellar radial velocities and a description 
of the it can be found elsewhere in the literature (Cochran \& Hatzes
1994; Butler et al. 1996). A description of the RV analysis 
techique as applied to
roAp stars can be found in Hatzes \& K\"urster (1992). 
Molecular iodine  has useful lines in the wavelength range of 5000--6300
{\AA} (14 spectral orders) which dictated the sub-framing of the CCD detector.

The pulsational amplitude ranged from a low  of 7 {\ms} in the spectral
order centred on 5928\,{\AA} to  a high of 57 {\ms} in orders centred on
5728\,{\AA} and 5456\,{\AA}. It is also clear from Figure~\ref{fig:dft}
that the
pulsational amplitude is not a simple function of wavelength. (We
caution the reader that the spectral order whose central wavelength is
5928 {\AA} is contaminated by a plethora of atmospheric water lines
and these may affect the RV amplitude derived from  the full spectral
order.)

\subsection{RV Amplitude and Phase Variations of Individual Lines}

\begin{figure*}
\epsfxsize=14truecm
\epsffile{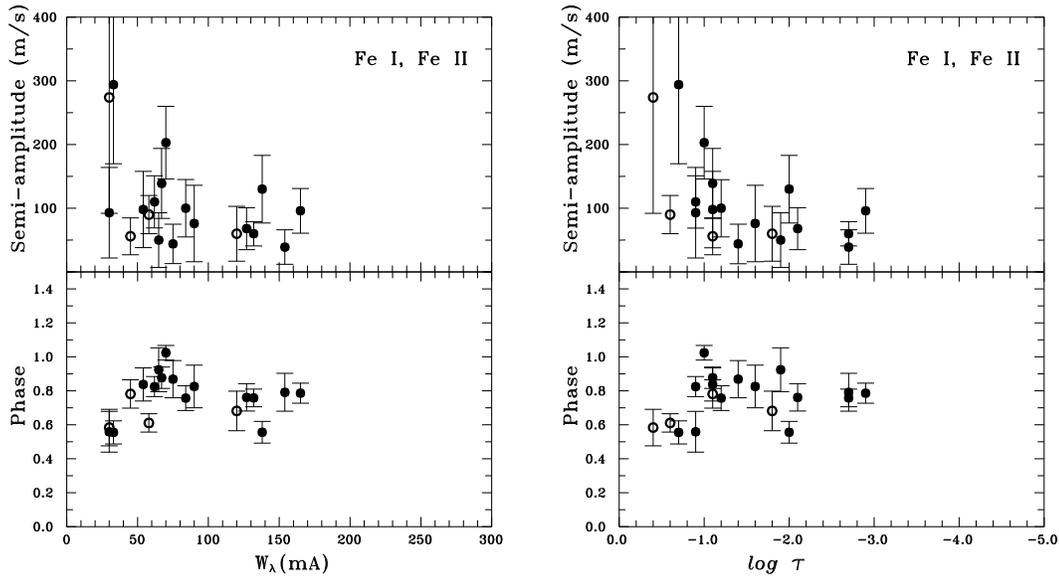}
\caption{The pulsational semi-amplitudes (top panels) and phases 
as a fractional pulsation period (lower panels)
versus line equivalent  width (left),  and 
optical depths (right) for Fe lines in 33 Lib.
The open circles represent the Fe II while the  dots are  for Fe I lines.
Phases in this and all subsequent figures are in units of
fraction of a pulsation
period.
\label{fig:fewap}}
\end{figure*}

\begin{figure*}
\epsfxsize=14truecm
\epsffile{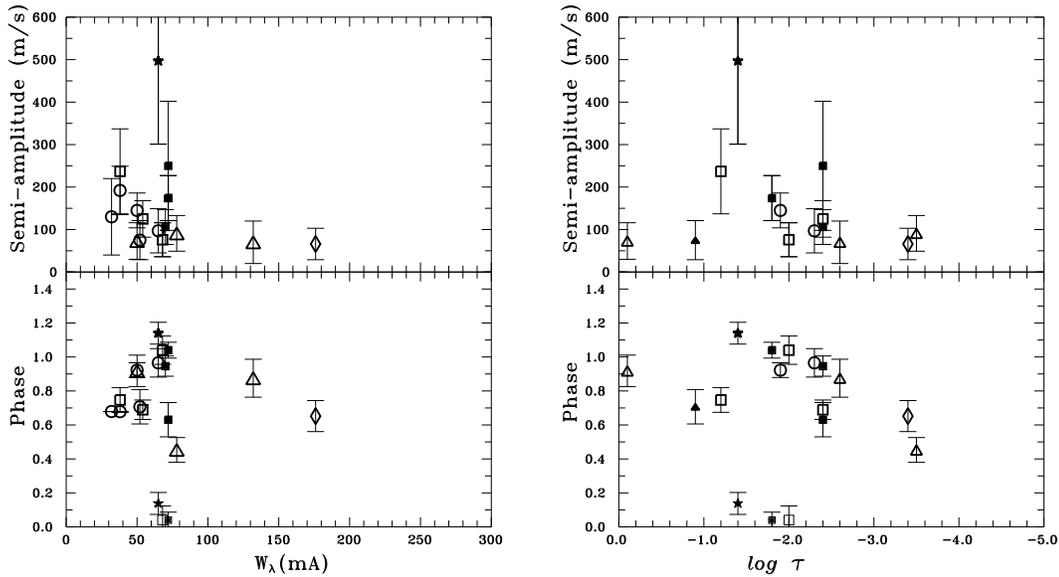}
\caption{The same as in Fig.\,\ref{fig:fewap} but for  Gd II (filled squares),
Er II (stars), La (open squares), Ni I (filled triangles), Ca I (open diamonds),
Ce II (open circles) and Cr II (open triangles) lines. The 3 lower
points near phase 0 have been replotted with phase $\phi$ = $\phi$ + 1.
\label{fig:rest}}
\end{figure*}

\begin{figure*}
\epsfxsize=14truecm
\epsffile{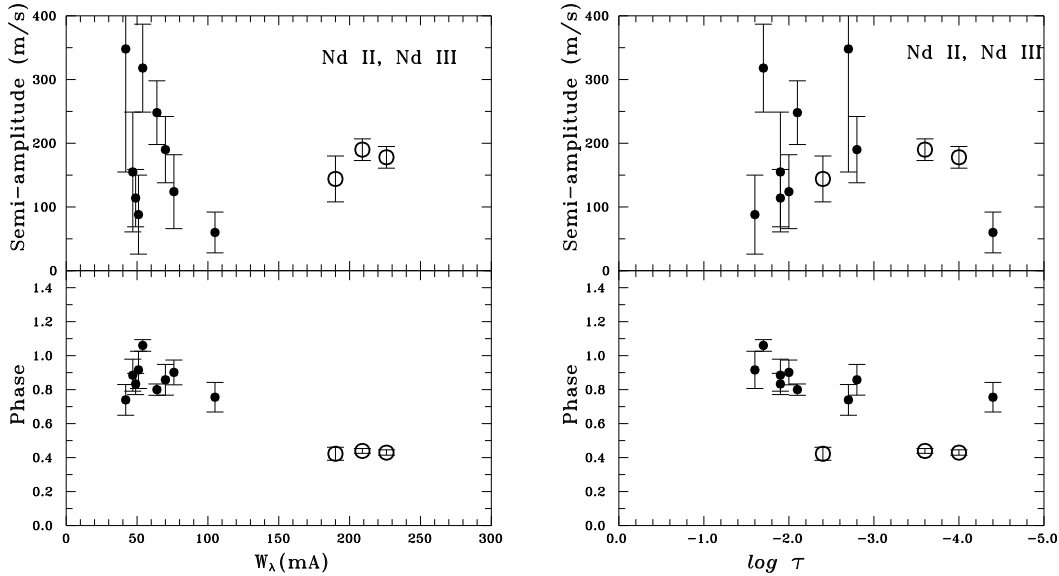}
\caption{The Nd II (dots) and Nd III (open circles) pulsational
amplitudes (top left and right panels) and phase (bottom left and right panels)
versus equivalent widths (left panels) and approximate optical depths (right panels) in
33 Lib. 
\label{fig:ndewt}}
\end{figure*} 

\begin{figure*}
\epsfxsize=14truecm
\epsffile{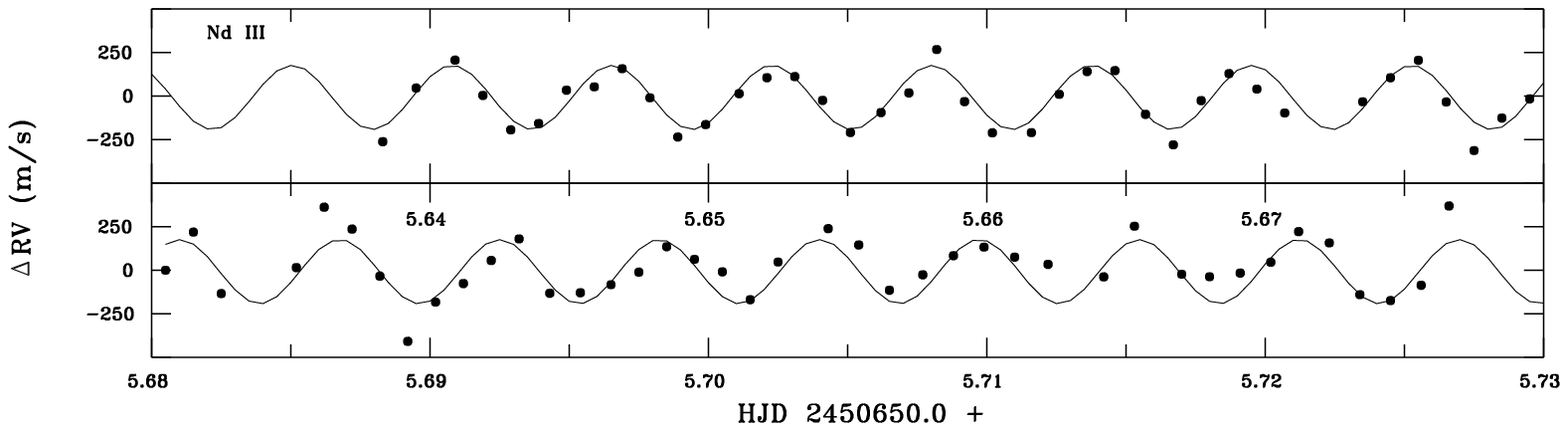}
\caption{The average radial velocity measurements
for Nd III  5102.42~{\AA} and Nd III 5294.10~{\AA} (dots). The
 solid line is sine-wave fit with pulsation
period 8.272 min. \label{fig:nd3vr}} \end{figure*}

To understand  the RV pulsational behavior of 33 Lib as a function of
atomic species and line strength we performed an RV analysis on
individual spectral lines. The rich spectrum of iodine absorption lines
enabled us to do so. Radial velocities were determined over a spectral
region containing only a single line. Care was taken to select only
unblended lines, although in some cases our analysis was performed on
slightly blended lines for which the major contribution to the
equivalent width came from one feature. Because of the increased noise
level in the RV measurements of individual lines, it was often the case
that  the dominant peak in the Fourier transform was not always
coincident with the photometric frequency. In these instances the
statistical significance of the highest peak was assessed using the
procedure outlined in Scargle (1982) and Horne \& Baliunas (1986).  Of
those peaks which did not coincide with the photometric frequency, none
were found to be statistically significant (the false alarm probability
was always greater than 10\%). Thus our line-by-line analysis did not
show any conclusive evidence for the presence of an RV frequency  that
differed from the dominant photometric frequency ($\nu_1$=2.0148 mHz). Table 1
lists the least squares RV amplitude and relative phases 
(as a fractional pulsation period) of maximum
RV amplitude. The relative pulsation phase was calculated using the time of
RV maximum from a least squares fit that  was phased with respect to the
time of the first observation ($JD$ = 2,450,655.6383). In this
convention phase 0.88 is near RV maximum. There appears to be
significant phase variations for all spectral features that were
measured along with  a tentative line identification when possible. In
our paper line identification were carried out based on line-lists
of the Vienna Atomic Line Data base (VALD) (Piskunov et al. 1995),
line-list for roAp star HD 122970 (Ryabchikova et al. 2000)
and a line-list for Przybylski's star (Cowley \& Mathys 1998). The blends are
marked in Table 1 by asterix and their wavelengths and identifications
are listed below the primary line.

	A major goal of this line-by-line analysis is to 
develop an approach for  deriving both the horizontal
and vertical structure to the pulsations. The
horizontal effects can be studied if we have a knowledge of the surface
distribution of elements on the star. It is well known from Doppler
imagery that non-oscillating Ap stars have elements that are
inhomogeneously distributed over the stellar surface (Hatzes 1991; Rice
\& Wehlau 1991,1994).  If an element is concentrated in a localized
spot, then the cancellation effect for non-radial modes will be smaller
for the RV variations from this element. Such a ``periodic spatial filter'' can be used to
derive additional information about the ($l,m$) quantum numbers
of the  nonradial modes (Mkrtichian 1994). Unfortunately, no Doppler image exists for 33 Lib due to long
rotation period and there has been no rotational modulation of spectral
features reported for this star. Thus we do not even have  crude
knowledge as to how elements are distributed on this star.

    The study of any vertical structure to the pulsations requires that
the acoustic amplitude and phases of a spectral lines be linked to a
geometrical depth in the atmosphere. 
However, unlike for normal stars the effective optical depth of formation of a 
spectral line cannot be easily linked to a geometric
depth in the atmosphere of a roAp stars.
The inhomogeneous distribution of elements both horizontally  and possibly
vertically, as well as the presence of strong magnetic fields
complicates the relationship  between optical and
geometric depths in roAp stars.
The optical depth calculated under simple assumptions (e.g. no magnetic
fields, no vertical stratification, etc.) may be considered
as a very approximate depth scale for the study of vertical atmospheric
cross-sections.

As a first approach 
we decided  to link the equivalent widths of the Fe spectral
lines, $W_{\lambda}$, with an optical depth, {\lgt}, of the line formation and
use this scale for estimating the  phase and amplitude distributions
throughout the atmosphere.
For this approximate {\lgt} scale  we assumed
that there is no vertical stratification of Fe and other elements in the
atmosphere of 33 Lib. This stratification may indeed be present as has
been suspected for several elements in Ap stars (Babel \& Lanz 1992;
Ryabchikova et al. 2002). Lacking any knowledge of
such stratification in 33 Lib this approach seems reasonable at least
for Fe since this element is non-peculiar and has approximately a solar abundance in
roAp stars (Ryabchikova et al. 2000). This {\lgt} scale for Fe will
be used as the standard scale for comparison with acoustic
cross-sections using calculated {\lgt} scales of other elements.

The tentative abundance determinations for all unblended lines of elements
were done  using  a version of
Kurucz's program WIDTH9 modified by V.V. Tsymbal (Tavrian National University)
and A.V. Yushchenko (Odessa National University) so that it could
use input files in the VALD format.
The  optical depths $\tau$ of spectral line formation were calculated
using the SYNTH-NEW code written by V.V. Tsymbal.

In cases where the atomic data were not present the VALD database (true
mostly for rare earth elements), the  optical depths were not
calculated. For the model atmosphere of 33 Lib we used $T_{eff}$ = 7350
K, $log$\,$g$ = 4.4 (Ryabchikova et al. 1999).
The microturbulent
velocity of $\zeta$=2.0 $km\,s^{-1}$ was derived by us using the unblended Fe I and
Fe II lines.

%

   Figure~\ref{fig:few}  shows the resulting  $W_{\lambda}$--{\lgt}
dependence for Fe I (dots) and Fe II (open circles) lines in the range
of $W_{\lambda}$=30--165 (m{\AA}). The dashed
line shows the linear fit expressed as {\lgt}= $-$0.2132 
$-$ 0.01560$W_{\lambda}$(m{\AA}).
As expected the 
weaker lines are formed on
average deeper in atmosphere than stronger ones.

\subsubsection{Fe lines}

Figure~\ref{fig:fewap}  shows the amplitude, $K$ (top panels), and
phase, $\phi$ (bottom panels), variation for the Fe lines. 
(In this and all subsequent figures phases are in units of fraction
of a pulsational period.) The panels
are divided between those showing $W_{\lambda}$ (left) and {\lgt}
(right) as the ordinates. The equivalent width is a measured quantity,
thus the $K$--$W_{\lambda}$, $\phi$--$W_{\lambda}$ plots are model
independent. On the other hand the calculated optical depth is dependent
on the assumptions used in the model atmosphere.
Because roAp
stars have large magnetic fields, peculiar abundances, and inhomogeneous
temperature and abundance profiles, both horizontally and vertically, the
uncertainties in the {\lgt} calculation are larger 
than for normal stars. The $W_{\lambda}$ --  {\lgt} mapping could change
drastically with improved modeling.  For this reason we show figures using
both $W_{\lambda}$ and {\lgt} as the ordinate.

\begin{figure}
\epsfxsize=8.5truecm
\epsffile{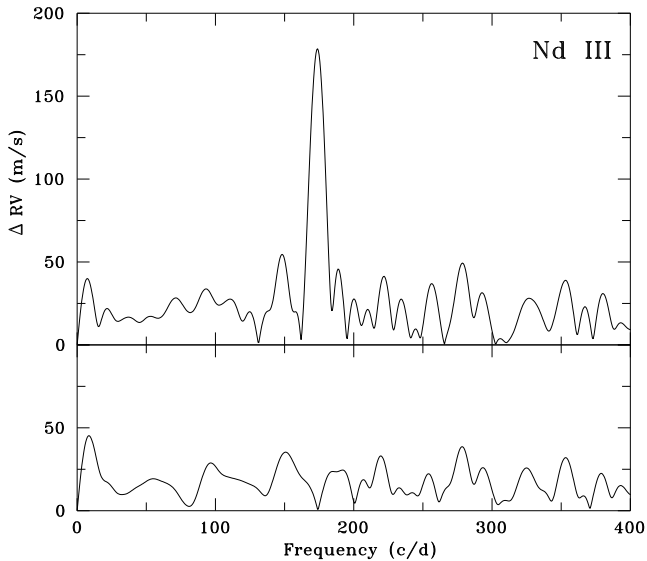}
\caption{The discrete Fourier transform of  the RVs for the Nd III lines
shown in Figure~\ref{fig:nd3vr}. The highest peak
corresponds to pulsation frequency 174.081 c/d. The  lower panel represents
the amplitude spectrum of the RV residuals after pre-whitening of this signal.
\label{fig:vrf}}
\end{figure}

\begin{figure}
\epsfxsize=8.5truecm
\epsffile{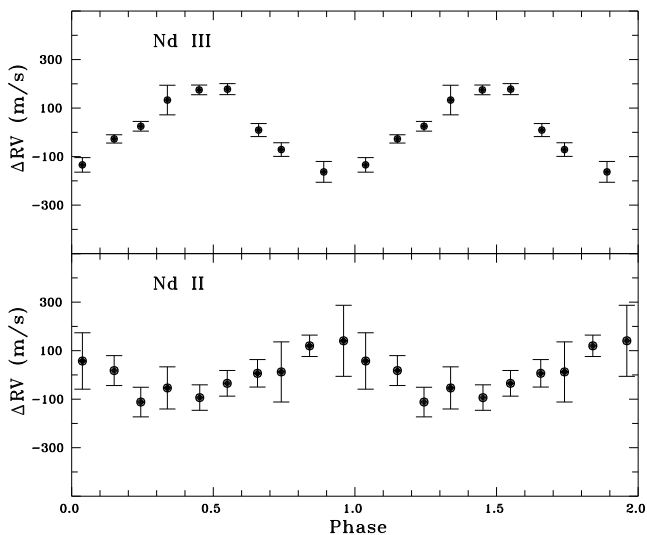}
\caption{The mean radial velocity
of Nd III (top panel) and Nd II (bottom  panel) lines 
phased to a  pulsational period 8.272 minutes.
\label{fig:nd23}}
\end{figure}

There appears to be a weak increase in the pulsation amplitude
with decreasing $W_{\lambda}$ and increasing {\lgt} for Fe lines.  However, the weaker lines
also tend to have much larger errors in the amplitude measurement.
This trend of increasing amplitude with decreasing line strength may
just be an artifact of the larger measurement error for weak lines.
Data of better quality is needed to establish
this with certainty.

The  $\phi$ -- $W_{\lambda}$, $\phi$ -- {\lgt} diagrams for Fe lines do
show, however, strong evidence for phase variability in the form of a 
phase jump of $\Delta\phi$ $\approx$ 0.3--0.5 at {\lgt} $\approx$ $-$0.9.

\subsubsection{Rare earths and other  elements}

Rare earth elements are known to show the largest abundance
anomalies and peculiarities in the atmospheres of Ap stars.   Due to
strong blending in 33 Lib the number of available unblended or partially
blended rare earth elements (REE) lines was limited for our analysis
(see Table 1.) Figure ~\ref{fig:rest} shows the amplitude (left and
right upper panels) and phase (left and right bottom panels) variations
respectively for Gd~II, Er~II, La~II, Ce~II REE elements,  and Ni I,
Cr II and Ca I lines. Phase is usually displayed
with values between 0 and 1 and with no information on the pulsation cycle.
This can influence how one interprets the phase diagram. The lower 
points (clustered around $\phi$ $\approx$ 0) have been replotted at
$\phi$ $\approx$ 1. Doing so makes these points more consistent with the
trends shown by the phase of other spectral lines.

The Er~II 5454.27 {\AA} line  shows the largest pulsation amplitude of 497
m/s (S/N=2.54) among measured lines. Again, similar to Fe lines,
large errors in the amplitude determinations and small number of
lines of REE, Ni, Cr II and Ca I lines  prevents us from
extracting reliable information about the amplitude dependence with
equivalent width.

The spectral features  of
Gd~II, Er~II, La~II, Ce~II, Ni I, Ca I,
and Cr II show a decrease in the pulsational phase with
decreasing optical depth for  {\lgt} $<$ $-$ 1.4.
There is also the distinct  phase jump of  $\Delta\phi$ $\simeq$ 0.4
at {\lgt} $\approx$ $-$1.3 that may be related to a similar phase jump at
{\lgt} $\approx$ $-$0.9 for the Fe lines. If both phase jumps are located
at the same geometrical depth in the stellar atmosphere (most likely) then the
difference $\Delta${\lgt} $\approx$ $-$0.4 in optical depths may be
attributed to the distortion  of {\lgt} scale due to different spatial
(horizontal and vertical) abundance distributions of REE.

We identified a sufficient number of Nd lines to be able to
separate these between singly and doubly ionized species.
The resulting amplitude and phase dependence versus the $W_{\lambda}$
and {\lgt} for Nd~II (dots) and Nd~III (open circles) lines are given
respectively in the left and right panels in Figure~\ref{fig:ndewt}.
The $K$--$W_{\lambda}$ relationship shows an  increase in the
 pulsation
amplitude with decreasing $W_{\lambda}$ which is not evident
in the $K$--{\lgt} diagrams. Again, large
rms amplitude scatter for weak Nd~II line
prevents us from drawing reliable conclusions about the amplitude changes
versus the depth in atmosphere.
The phase dependence for Nd II lines  shows a slight decrease in the 
pulsation phase for optical depths smaller than 
{\lgt} = $-$1.6.

Three strong Nd III lines 5294.10~{\AA}, 5102.42~{\AA} (partially
blended with Nd II 5102.39~{\AA} line) and 6145.07~{\AA} lines provided
RV measurements of exceptional quality. All lines had comparable RV 
comparable amplitudes  and their phases were consistent to within 
$\pm$0.005 of pulsation period. The relatively small scatter in the
RV measurements of these lines may be due to the fact that
they  are formed, on average, in the same  atmospheric layers.

The averaged RV measurements for Nd III 5102.42 {\AA} and 5294.1 {\AA}
lines as a function of time are shown in Figure~\ref{fig:nd3vr}.
Figure~\ref{fig:vrf} shows the DFT amplitude spectrum of these data (top
panel). The tallest peak with an  amplitude of 184.3 \ms corresponds to
known photometric frequency $\nu_{1}$=174.081 c\,d$^{-1}$. The 
amplitude spectrum of the 
residuals obtained after pre-whitening the 174.081 c\,d$^{-1}$ signal 
is shown at
bottom panel in Figure~\ref{fig:vrf} and it does not show any peaks
above the noise peaks level of 38 {\ms}.
The second harmonic of $\nu _{1}$
$\nu _{2}$=348.16 c\,d$^{-1}$ which is found in the photometric data (Kurtz
1991) with an amplitude ratio A($\nu_{2}$)/A($\nu _{1}$) = 0.13 is not
present in the Nd III data above the noise level. If we assume the same
ratio of amplitudes of the first and second harmonic for the RV, then
the expected RV amplitude of the second harmonic should be about 24 {\ms}
-- below our detection limits using radial velocities
measured from the strong
Nd III lines.

The striking feature about Figure~\ref{fig:ndewt} is that the 
mean phase of the Nd III lines differs by about 180$^\circ$ from the mean
phase of the Nd II lines.   This is shown more clearly in Figure~\ref{fig:nd23}.
The bottom panel shows the averaged RV
measurements of the four weak Nd~II 5456.55 {\AA}, 5614.28 {\AA},
5625.73 {\AA} and 5761.69 {\AA} lines phased to photometric period.
The mean pulsation $K$-amplitude of Nd II
lines is K=111 $\pm$ 45  m\,s$^{-1}$. The top
panel in Figure~\ref{fig:nd23} shows the phased mean radial velocities
measured of the two strong 5294.10 {\AA}  and 5102.42 {\AA}  Nd~III
lines. The mean $K$-amplitude for the Nd III
lines is 175 $\pm$ 10 m\,s$^{-1}$. Clearly, the Nd II and Nd III lines are
pulsating almost 180{\deg} out-of-phase ($\Delta$$\phi$ = 0.46) with
respect to each other.

The lower right panel of  Figure~\ref{fig:ndewt}  shows an obvious
discrepancy in the {\lgt} of Nd II and Nd III features. Both ionized
species have a comparable range of optical depth, yet their pulsation 
phases are antiphase. This cannot be the case if both species were formed
at the same depth in the atmosphere. This discrepancy can be resolved
if the strong Nd III lines were in fact formed much higher in the stellar
atmosphere and their calculated (under simplified assumptions about
homogeneous vertical elemental distribution) approximate optical depths
are essentially overestimated.

Normally, we expect the second ionized species to form in the deep
atmospheric layers where the temperature is high or in the outer layers
where  the pressure is low. The $\phi$ -- $W_\lambda$ diagram
(lower left of Figure~\ref{fig:ndewt}) argues for the former.
However, if Nd III lines are indeed formed at much higher
layers then this could explain the differences in pulsational phase
between Nd II and Nd III and points to the existence of a
standing acoustic wave with a node in the upper atmosphere. 
We could not
determine accurately the position of this node because the acoustic
cross-sections are not well sampled by spectral lines at small optical depths;
we can only estimate the position of this pulsation node as somewhere above
{\lgt} = $-$4.5 and that Nd III lines are formed above this node at
essentially small {\lgt}. Support for this result also comes from
Ryabchikova et al. (2002)  who found by a trial-and-error approach in LTE
line-profiles synthesis  that Nd is strongly stratified in the 
atmosphere of
$\gamma$ Equ and it has a thin overabundant  (by more than 6 dex) layer
above a height {\lgt} $\approx$ $-$8.0. If a similar vertical stratification
of Nd occurs in the atmosphere of 33 Lib, then this
would alter the {\lgt} -- $W_{\lambda}$
relationship.

\subsubsection{Mean Phase and Amplitude Distributions}

\begin{figure}
\epsfxsize=8.5truecm
\epsffile{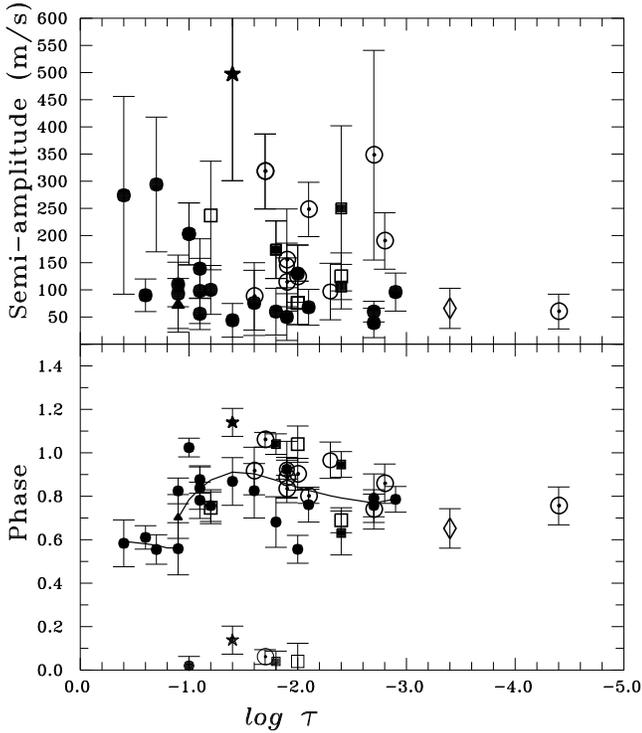}
\caption{The combined acoustic $K$-{\lgt} (upper panel) and  $\phi$--{\lgt}
(bottom panel) cross-sections of atmosphere of 33 Lib for unblended
Fe I, Fe II (dots), Gd II (filled squares), La II (open squares),
Ce II (open circles), Er II (stars),  Ni I (filled triangle), Ca I (diamonds)
 and  Nd II (sun symbols).
The solid line in the bottom panel shows 
polynomial fits to the phase versus  {\lgt}.
A first order fit was used for  {\lgt} $>$ $-$0.9 and 
ninth order for  {\lgt} $<$ $-1.0$.
A phase jump at {\lgt} $\simeq$$-$0.9 is evident.
\label{fig:apht}} \end{figure}

Figure \ref{fig:apht} shows  the amplitude (top panel) and phase (bottom
panel) versus optical depths for unblended Fe, Cr, Gd, Er, La,
Ce, V, Ca, Ti spectral lines. This
{\lgt}-$\phi$ diagram for the majority of unblended lines also shows 
evidence for the existence of phase variability in acoustic cross-sections.
In particular, in addition to the 
phase jump at {\lgt} $\approx$ $-$0.9 there is also evidence of slow
phase changes above {\lgt} $<$$-$0.9 and below {\lgt} $>$ $-$3.0.

\begin{figure}
\epsfxsize=6.75truecm
\epsffile{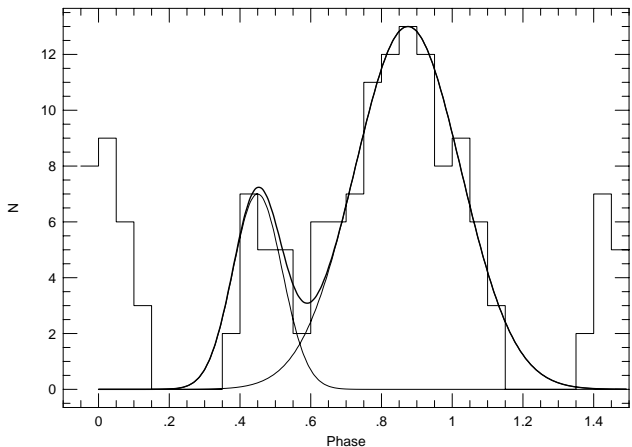}
\caption{Histogram of the distribution of pulsational phases 
(in fraction of a pulsation period) for all
spectral lines. The thin lines represent Gaussian fits to the two
distributions with $\sigma$ = 0.15 for the large Gaussian and
$\sigma$ = 0.07 for the small Gaussian. The thick line represents
the sum of the two component Gaussians. The approximate position
of acoustic node is expected at relative $\phi$$\simeq$0.6.
\label{fig:histo}}
\end{figure}

We also checked  whether phase clustering  which was found in 
Fe and REE elements lines was also present in  all spectral lines and blends of
33 Lib.  The histogram of relative
pulsational phases for all lines is shown in Figure~\ref{fig:histo}
(0.05 phase bins) and this shows a clear bi-modal distribution. Most spectral
lines  are centred on $\phi$ = 0.88 with a spread that is consistent
with a Gaussian distribution of  half width, $w$ = 0.15. Nineteen of
the spectral lines we examined had a pulsation phase near 0.425 or
164{\deg} apart from the phase of most other lines. These also appear to
have a Gaussian distribution  with half-width $w$ = 0.07. The individual
Gaussian distributions are shown as thin lines in the figure while the
thick line represents the sum of the two component distributions.
The phase $\phi \approx$ 0.6 corresponds to borderline between the two
distributions. This bi-modal distribution for 33 Lib is
consistent with the Baldry et al. (1998) result for $\alpha$ Cir.
Interestingly, the fraction of our spectral lines having a phase nearly
180{\deg} from the mean is about 17\%,  near the 15\% fraction of
the bands showing a 180{\deg} phase shift in $\alpha$ Cir found by
Baldry et al (1998). The phases of the individual lines have a mean
standard deviation of 0.09 which is consistent with the width of the
smaller Gaussian. However, the standard deviation of the distribution of
phases centred on 0.88 is larger than this by nearly a factor of two.
This suggests  there is  additional variations that cannot be accounted
for by the errors of the phase measurements.

The profile of the phase distribution on opposite sides of phase jump
{\lgt} $\approx$ $-$0.9 (which is interpreted in {\S}5 as the
second acoustic node) seems to contradict the pure standing wave picture which requires
a  sharp 180{\deg} phase jump and non-variable phase on opposite sides
of acoustic node.

The phase diagrams of Figures~\ref{fig:fewap}, ~\ref{fig:rest},
and ~\ref{fig:apht} appear to be a superposition of  a step
function (i.e. discontinuous jump in phase) along with a linear function
of decreasing phase with {\lgt}. That is to say, the phase after
(and possibly before) the phase jump at {\lgt} $\approx$
$-$1 is not constant, rather it decreases linearly as one moves outward
in the atmosphere (smaller values of {\lgt}).
The running wave component and linear decrease in phase outward in 
the atmosphere
can  qualitatively
explain the  non-equality  to 180{\deg} in the phase  spacing between
of centers of distributions of two gaussians in Figure~\ref{fig:histo}.

To establish reliably the presence of running wave component in the atmosphere
requires more accurate determinations of the phases than presented in this study.

\subsubsection{Effect of the magnetic field}

The strong stellar magnetic fields can influence the pulsations in roAp
and  are believed  to cause the high-overtone oblique pulsations with
the axis of symmetry coincident with (Kurtz 1990) or  close to 
(Bigot \& Dziembowski 2002) the magnetic axis . The
magnetic fields can also influence spectral lines through magnetic
Zeeman splitting and lead to line broadening and intensification effects
which can influence the depth of formation. It was thus natural to
investigate whether  the pulsational phase or amplitude  of
spectral lines correlated with their magnetic sensitivity. As a
parameter of magnetic sensitivity the effective Lande factors $g_{eff}$
of spectral line transitions were taken from VALD data base compilation.

\begin{figure*}
\epsfxsize=14truecm
\epsffile{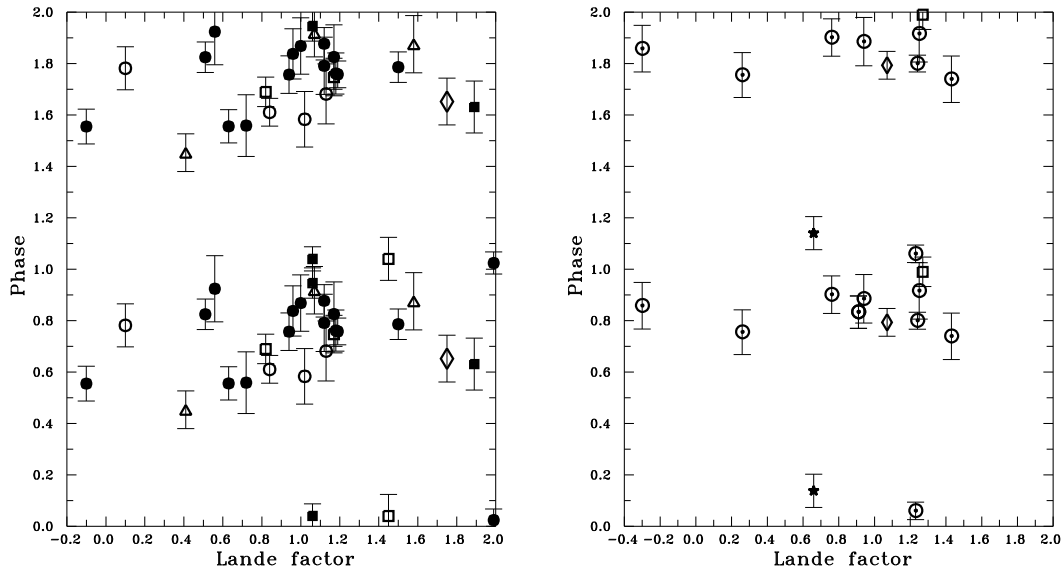}
\caption{ The left panel: the pulsation phase  (fraction
of a period) versus the effective Lande g-factor for
Fe I and Fe II (filled circles),  Cr II (open triangles), Gd II (filled
squares), La II (open squares) and  Ca I (open diamonds).
The right panel: the same for Nd II (sun symbols), Dy II (open diamond),
Er II (stars),  and V II (open square).
\label{fig:lande}}
\end{figure*}

We found a puzzling, yet  well defined  dependence of the pulsation phase with
effective Lande g-factors for some species.
Figure~\ref{fig:lande} shows the pulsational phase versus
effective Lande factors $g_{eff}$ dependencies for Fe I and Fe II, Gd II,
La II, Cr II and Ca I lines. The phases are shown over 2 cycles
(with the same data points)
for better presentation of phase variability.

The relative pulsation phase rises linearly for lines with 
$g_{eff}$ $>$ 0.6 reaching a maximum at $g_{eff}$ $\approx$  1.05
The phase then decreases for 1.05 $<$ $g_{eff}$$<$ 1.2.
Over these intervals the phase variations form 
what appears to form an inverted `V'-shape. There  is a 
discontinuity in the pulsation phase for $g_{eff}$ $>$ 1.4. The
phase variations for these higher $g_{eff}$-lines decrease linearly with 
phase,
but are shifted by about 0.5 in phase with respect to the linear trend
seen for 1.1 $<$ $g_{eff}$$<$ 1.2.
However, these points can form a continuous phase variation to the
lower $g_{eff}$ lines if one considers that the spectral
lines with $g_{eff}$ $>$ 1.4 are actually shifted by one complete pulsational
cycle.

The right panel
of Figure~\ref{fig:lande} shows the same diagram as the left panel,  but
for Nd II , Dy II , Er II and V II lines. These do not show  any well
defined dependence between phase and effective Lande factors. This discrepancy
could be explained by different spatial distributions of these elements
in comparison with Fe, Cr, Gd, La and Ca lines.
Assuming the presence of a vertical acoustic wave in the atmosphere the
pulsation phases variations versus the $g_{eff}$ for Fe, Cr, Ca, Gd and
La lines may be considered as due to the strong influence of magnetic fields on the
depth of formation of spectral lines.

According to the acoustic cross-sections of the atmosphere of 33~Lib
the vertical scale of the acoustic  wave is small and is comparable to the
thickness of the line forming region of the stellar atmosphere.
The left panel of Figure~\ref{fig:lande} suggests that magnetic intensification effects
introduce changes in the line formation depth  that is comparable to the
vertical wavelength of the acoustic wave. Thus over this vertical distance there
are significant phase changes.

\begin{figure}
\epsfxsize=8.5truecm
\epsffile{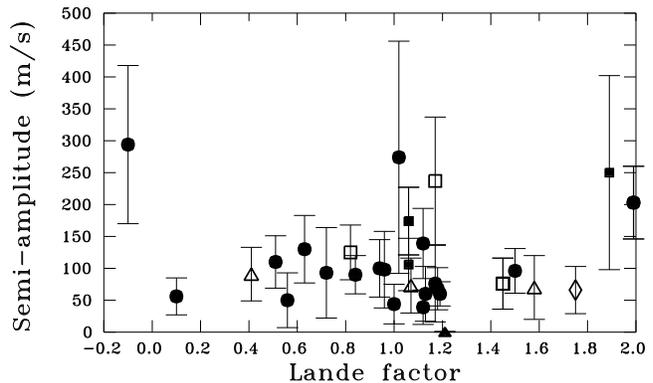}
\caption{ The  pulsation semi-amplitude  versus effective Lande g-factor for
Fe I and Fe II (filled circles),  Cr II (open triangles), Gd II (filled
squares), La II (open squares),  Ca I (open diamonds).
and Ni I (filled triangles).
\label{fig:landa}}
\end{figure}

However, the $g_{eff}$--{K} diagram in Figure\,\ref{fig:landa} does not
show a correlation between pulsation amplitudes versus the effective Lande
g-factors.  A $g_{eff}$--{\lgt} relationship determined with the
inclusion of magnetic effects on the line formation should establish
whether the phase changes are due to
variability of effective depth of line formation induced by
the magnetic sensitivity of lines.

The approximately  2 hour duration of our RV data does not allow us to
probe the origin of $g_{eff}$--${\phi}$ dependence 
in 33~Lib.
Future investigations with more accurate phase and
amplitudes determinations and a better
description  of atmospheric elemental stratifications and magnetic
intensification of lines are needed to get a better
understanding of this effect.

\section{Discussion and conclusions}

Our pulsational phase data indicate the presence at least one pulsation
node in the upper atmosphere, and probably a second one in the deeper
atmospheric layers of 33 Lib.
The strongest evidence of upper node comes from the RV
variations for Nd II and Nd III which  are almost 180$^\circ$
out-of-phase with each other (Fig.~\ref{fig:nd23}). Since Nd II and Nd
III should be formed in different atmospheric layers these data
provide convincing evidence for a upper atmospheric  pulsational node.
We estimate that the node is located above {\lgt} $<$ $-$4.5.

A second phase jump in the lower atmosphere at {\lgt}  $\approx$
$-$0.9 is seen in the
 weak Fe spectral lines
(Fig.~\ref{fig:fewap}) and this may be associated with another node.
The weak Fe lines with {\lgt}  $>$ $-$0.9 oscillate out-of-phase ($\Delta
\phi \approx$ 0.3-0.5) with those of stronger lines formed at {\lgt} $<$
$-$0.9.
The mean relative phase of these four weak Fe I and Fe II lines is 
$\phi$ $\simeq$ 0.6, or about the same value that defines
the border between the two distributions in  Fig.~\ref{fig:histo}.
So, we suspect another, lower  node below
or at {\lgt} $\approx$ $-$0.9 closer to the continuum formation level.

The hypothesis of two acoustic nodes in the atmosphere of 33 Lib
can qualitatively explain the $\approx$ 17\% fraction of spectral lines
and blends showing out-of-phase oscillations in Figure~\ref{fig:histo}.
According to this hypothesis the upper node is located somewhere above
{\lgt} $\approx$ $-$4.5 and the second one below, or at
{\lgt} $\approx$ $-$0.9, closer to the continuum formation level.
In this case the layers where the majority of spectral lines are
formed  are pulsating with relative phases centred  around $\phi$=0.88
and are 
between the two nodes. The other spectral lines
that contribute to the smaller Gaussian distribution of 
phases centred  on $\phi$=0.425 are
formed in two, out-of-phase pulsating layers: either below the lower node
({\lgt} $>$ $-$0.9), where the weaker lines are formed  or above 
the upper node  ({\lgt} $<$ $-$4.5) where the few Nd
III lines and some strong blends are formed.

	Theoretical work of  Gautschy et al. (1998) supports our hypothesis
about two radial pulsation nodes  in the atmosphere of an roAp star.
Gautschy et al.  have calculated a sequence of models for roAp star with
a temperature inversion in the upper atmospheric layers.  These models
predict two displacement nodes in the stellar atmosphere: one close to the
continuum formation  $\tau$ $\approx$ 2/3 and another at superficial
layers (see Fig. 13 in their paper). The location of nodes in atmosphere
is model and mode dependent, so the presence of two nodes in the line forming
region is in principle theoretically possible.


%

	The inverted 'V'-shaped variations in pulsational phase with the effective
Lande g-factor
(Fig.~\ref{fig:lande}) in 33 Lib were unexpected.
Stift \& Leone (2003)
presented an in-depth investigation of the increase in equivalent width
of saturated lines under the influence of magnetic fields. Assuming
20 different Zeeman patterns (with the number of 
components ranging from 3 to 45) they calculated equivalent widths as a 
function of magnetic field strength and showed that the 
equivalent width can increase by up to factors of 10.
Thus, the magnetic field
may strongly affect the in-line absorption  and hence the effective
depth of formation and this could result in a effective Lande g-factor that is
correlated with pulsation phase.  This is consistent with phase changes
across the atmosphere where the thickness of the line-forming region 
is comparable
to the vertical wavelength of the acoustic wave, and that the influence
of the magnetic field through intensification effect also occurs over
this small region. Fig.~\ref{fig:lande} may thus reflect further
evidence of a pulsation phase variability  and  running wave component
in the stellar atmosphere. Further investigations of the effective Lande factor $g_{eff}$  versus pulsation
phase in other roAp stars may prove fruitful.

 Although RV variations have been measured for relatively few roAp
stars, these have already shown a rather complex behavior. In order to
interpret fully these variations it is essential to perform detailed
spectral analyses of these stars with a special emphasis on the
accurate determination of  the depth of formation for
each spectral line, accurate acoustic cross-sections of atmosphere
in lines of different chemical elements,  and the
study of their  variability over the
rotation period. This would yield knowledge about the acoustic profiles and
location of the radial node in different surface areas and with  respect to magnetic axis.

On another hand, the acoustic cross-sections can also be used to refine
the atmospheric models. The pulsation phase (and amplitude) distribution
across the atmosphere is a unique and true link with the geometrical
depth scale and could be used for atmospheric tomography. This tomograpic
procedure should include the semi-empirical adjustment and tuning of the
vertical and horizontal elemental distributions (the latter can be known
in advance, from Doppler imaging)   and physical conditions
in the model of atmosphere until the best solution for phase and
amplitude cross-sections versus geometric depths for different chemical
elements is attained (and hence true transformation scales between EW,
optical and geometrical depths obtained).
For example, in our simplified approach for a model atmosphere
of 33 Lib with homogeneous vertical distribution of Nd, the calculated
optical depths suggest that Nd II and Nd III are formed at comparable
optical depths which is inconsistent with the pulsational phases for Nd
III (Fig~\ref{fig:ndewt}). 

The restoration of vertical atmospheric structure and geometrical depths
of spectral line formation are only as accurate as the measured RV
pulsation phases and amplitudes of spectral lines formed in different
atmospheric layers. For the long duration (weeks) spectral observations
it is possible to increase the resolution in geometrical depths across
the line-forming atmosphere to $10^{-3}$ or better.

Moreover,  as the
different p-modes having different oscillation frequencies are expected
to have different depths for acoustic nodes in atmosphere
(Gautschy et al., 1998), this will
lead to the different phase (and amplitude) profiles in atmospheric cross-sections
measured on the same spectral lines but on different oscillation
frequencies (i.e. different modes). This expected effect gives an
independent and additional tool to sounding the vertical acoustic
structure of atmosphere and study vertical chemical abundances in
multi-mode roAp stars using different cross-sections.
Such an accurate tomographic analysis is beyond the scope of the current
work and requires longer time series  for an mono- or  multi-mode roAp star,
RV measurements of  much larger number spectral lines,
and data taken with better signal-to-noise ratio. These  data can be used to glean valuable information on the
3-dimensional structure to the pulsations and the stellar atmosphere of
roAp stars.

\section{Acknowledgements}
	This research was supported by  grant UP2-317 by the
Civilian Research and Development Foundation (CRDF) during
1997-1999 and a grant by the
German Academic Exchange Service (DAAD) during 2002. We express our
sincerest thanks to T. Ryabchikova for making available to us
linelist for roAp star HD\,122970. DEM would like thank A. Gautschy for
useful comments on his models of roAp stars and  V.V. Tsymbal
and A.V. Yushchenko for helpful comments on spectral synthesis and
analysis codes. DEM acknowledges his work as part of research acitivity of the
Astrophysical Research Center for the Structure and Evolution of the Cosmos (ARCSEC)
 which is supported by the Korean Science and Engineering Foundation.

\clearpage

\begin{table}
\begin{center}
\begin{tabular}{llccccccc}
Measured $\lambda$ & Identification      &$W_{\lambda}$   & $K$ Amplitude & Error & Phase & Error&$log~\tau$&$g_{\rm eff}$ \\
             {{\AA}} & {\hskip 20pt} {{\AA}} & m{{\AA}}& (m/s)         & (m/s) &       &      &    &  \\
\hline
         &               &     &     &     &           &     &    &\\
 5037.81 & 5037.78 Ce II &  65 &  97 &  52 &.97  &.08&-2.3  &0.84\\
 5048.01*& 5048.05 La II &  38 & 237 & 100 &.75  &.07&  &\\
         & 5048.06 Ni II &     &     &     &       &     &  & \\
 5061.08*& 5061.06 Gd II &  45 &  86 &  74 &.51  &.13&  &\\
         & 5061.08 Ni II &     &     &     &       &       &  & \\
         & 5061.09 Fe II &     &     &     &       &       &  & \\

 5065.07*&  5065.01 Fe I & 126 & 61  &  47 &.45  &.12&   &\\
         &  5065.03 Fe II&     &     &     &       &       & &  \\
         &  5065.04 Cr I &     &     &     &       &       & &  \\
 5070.97 & 5071.01 Gd II&  72 & 250 & 152 & .63 &.10& -2.4&1.89\\
 5082.30 & 5082.34 Ni I  &  52 &  75 &  46 &.71  &.10&-0.9 &1.21 \\
 5092.29* & 5092.25 Ti I  &  76 & 184 &  97 &.98  &.08& &     \\
         & 5092.25 Mo I &     &     &     &       &       &   &\\
         & 5092.25 Gd II &     &     &     &       &       &   &\\
 5102.50*& 5102.42 Nd III& 209 & 190 &  17 &.44  &.01&-3.6 &0.47\\
         & 5102.39 Nd II &     &     &     &       &       &   &\\
 5105.21 & 5105.23 Nd II&  42 & 348  & 193 &.74 &.09&-2.7& 1.43\\

 5121.70*& 5121.64 Fe I &  60 & 112  &  43 & .92 &.06& &\\
         & 5121.56 Ni I  &     &    &     &       &       &   &\\
 5127.94 & 5127.87 Fe II &  58 &  90 &  30 & .61 &.05&-0.6 &0.84\\
 5133.78*& 5133.69 Fe I & 176 & 137  &  45 & .95 &.05&   &\\
         & 5133.83 Er II &     &     &     &       &       &  & \\
 5171.59* & 5171.60 Fe I & 125 & 91  &  30 & .72 &.06&  & \\
         &  5171.64 Fe II &     &     &    &       &     &  & \\
         &  5171.67 Fe I  &     &     &    &       &     &  & \\
 5182.59 & 5182.59 Nd II&  70 & 190 &  52  &.86  &.09&-0.291& -0.3\\
 5183.56*& 5183.42 La II& 279 &  45 &  31 &.90  &.11& &\\
         &  5183.60 Mg I  &     &     &    &       &       &  & \\
         &  5183.71 Ti II &     &     &    &       &       &  & \\
 5188.76* & 5188.68 Ti II & 240 &  64 &  21 &.90  &.05& &\\
          & 5188.90 Er II &     &     &     &       &       & &\\
 5196.51*& 5196.48 Cr I  &  90 & 141 &  75 &.56  &.09&  &\\
         & 5196.59 Cr I  &     &     &     &       &       &  & \\
         & 5196.592 Mn I &     &     &     &       &       &   &\\

 5197.64*& 5197.58 Fe I & 125 &  68 &  63 &.04   &.14& &\\
         &  5197.48 Fe II &     &     &    &          &     &   & \\
         &  5197.66 Dy II &     &     &    &          &     &   & \\

 5208.54*& 5208.42 Cr I & 284 &  46 &  21  & .07  &.07&  &\\
         & 5208.59 Fe I  &     &     &     &          &     & &   \\
 5223.25 & 5223.19 Fe I  &  62 & 110 &  41 &.82   &.06&-0.90 &0.51\\
 5228.37* &5228.37 Fe I  &  65 &  86 &  39 &.88   &.07& &\\
         &  5228.40 Fe I &     &     &    &          &     &  &  \\
         &  5228.42 Nd II&     &     &    &          &     &   & \\
 5232.92* & 5232.92 Ce II& 170 &  42 &  32 &.06  & .12  &\\
         &  5232.95 Fe I  &     &     &    &          &     & & \\

 5242.48 & 5242.49 Fe I  &  75 &  44 &  31 &.87  &.11 &-1.4 &1.0 \\
 5246.78 & 5246.77 Cr II & 132 &  70  &  50 &.88  &.11 &-2.6 &1.58\\
\hline
\end{tabular}
\caption{Radial Velocity Amplitudes. All phases are in
fraction of  a pulsational period. }
\end{center}
\end{table}
\clearpage

\setcounter{table}{0}
\begin{table}
\begin{center}
\begin{tabular}{llccccccc}
Measured $\lambda$ & Identification &$W_{\lambda}$ & $K$ Amplitude & Error & Phase  & Error&$log \tau$&$g_{eff}$ \\
{{\AA}} & {\hskip 20pt} {{\AA}} & m{{\AA}}& (m/s)  & (m/s)     &           &        &&\\
\hline
 5252.02*&5252.03 Ce I  & 115 & 126 &  68 &.37  & .08 &&\\
         &5252.02 Ti II  &     &     &     &       &       &      &\\
         &5252.10 Ti I   &     &     &     &       &       &      & \\
 5257.05*& 5257.02 Er II & 140 &  55 &  54 &.39  &.15  &&\\
         & 5256.94 Fe II &     &     &     &          &      &&   \\
 5261.66*& 5261.70 Ca I  & 130 &  22 &  21 &.80   &.16 & &\\
         & 5261.76 Cr I  &     &     &     &          &      &&   \\
 5269.51*& 5269.48 Nd II & 146 &  72 &  65 &.64   &.15 & &\\
         & 5269.52 Ce II &     &     &     &          &      &&   \\
         & 5269.54 Fe II &     &     &     &          &      &&   \\
 5273.39*&5273.37 Fe I   & 140 &  70 &  52 &.00   & .11&  &\\
         &5273.43 Nd II  &     &     &     &          &      & &  \\
 5278.20*& 5278.20 Fe II &  61 &  96 &  68 &.95  & .054& &\\
         & 5278.25 Cr I  &     &     &     &         &       & &  \\
 5281.78 & 5281.79 Fe I  & 127 &  68 &  33 &.76  & .08 &-2.1&1.18\\
 5290.81 & 5290.82 La II &  54 & 125 &  43 &.69  & .06 &-2.4 &0.82\\
 5294.10 & 5294.10 Nd III& 226 & 178 &  17 &.43  & .02 &-4.0 &0.9\\
 5307.23*& 5307.23 Ca II  & 143& 9   &  27 &0.79   & 0.47 &   &\\
         & 5307.27 Cr I   &     &     &     &         &       &   &\\
         & 5307.36 Fe I   &     &     &     &         &       &   &\\

 5316.65*& 5316.60 Nd II & 250 &  75 &  26 &.73  & .06 & &\\
         & 5316.62 Fe II &     &     &     &         &       & &   \\
         & 5316.78 Fe II &     &     &     &         &       &  &  \\
 5319.79 & 5319.82 Nd II & 105 &  60 &  32 &.76  & .09 & -4.4&0.26\\
 5324.17 & 5324.18 Fe I  & 165 &  96 &  35 &.79  & .06 &-2.9&1.5\\
 5325.56 & 5325.55 Fe II & 120 &  60 &  43 &.68  & .12 &-1.8 &1.13 \\
 5334.21*& 5334.23 Er II& 113 & 131  &  62 &.97  & .07 &  &\\
         & 5334.24 Sc II &     &     &     &         &       & &   \\
 5334.84*& 5334.87 Cr II &  78 &  91 &  42 &.45  & .07 &-3.5&0.41\\
         & 5334.87 Mn II &     &     &     &         &       & &   \\
 5362.86*& 5362.87 Fe II& 156 &  61  &  46 &.92  &1.17 & &\\
         & 5362.77 Co I  &     &     &     &         &       & &  \\
 5364.87 & 5364.87 Fe I & 138 & 130  &  53 &.56  & .06 &-2.0 &0.63 \\
 5373.01 & 5373.01 Tm II &  50 &  93 &  64 &.50  & .11 &  &\\
 5373.70*& 5373.71 Fe I &  75 & 117  &  92 &.96  & .12 & &\\
         & 5373.72 Cr I &     &      &     &       &         && \\
 5379.19 & uncl         &  42 & 182 &  88  &.83  & .08 & &\\
 5383.37& 5383.37 Fe I  & 154 &  39 &  27 &.79  & .11 &-2.7  &1.12\\
\hline
\end{tabular}
\caption{Radial Velocity Amplitudes (cont.)}
\end{center}
\end{table}
\clearpage

\setcounter{table}{0}
\begin{table}
\begin{center}
\begin{tabular}{llccccccc}
Measured $\lambda$ & Identification &$W_{\lambda}$ & $K$ Amplitude & Error & Phase & Error&$log \tau$ &$g_{eff}$\\
{{\AA}} & {\hskip 20pt} {{\AA}} & m{{\AA}}& (m/s)  & (m/s)     &           &       & &\\
\hline
 5395.20 & 5395.22 Fe I  &  33 & 294 & 124 &.55  &   .07 & -0.7 &-0.1 \\
 5395.82*& 5395.75 Cr II &  32 & 139 &  36 &.89  &   .04& &\\
         & 5395.90 Er II &     &     &     &         &        &&   \\
         & 5395.63 Cr II &     &     &     &         &        & &  \\
         & 5395.86 Fe II &     &     &     &         &        &  & \\
 5402.78*& 5402.76 Eu I  &  56 &  92 &  41 &.78  &   .08&&  \\
         & 5402.77 Y  II  &     &     &     &         &        & &  \\
 5407.59*& 5407.60 Cr II & 104 & 140 &  45 &.87  &   .05& &\\
         & 5407.42 Mn I  &     &     &     &         &        & &  \\
         & 5407.48 Fe I  &     &     &     &         &        &  & \\
 5429.74*& 5429.50 Fe I  & 312 &  35 &  39 &.41  &   .17& &\\
         & 5429.70 Fe I  &     &     &     &         &        & &  \\
         & 5429.83 Fe I  &     &     &     &         &        &  & \\
 5439.72 & 5439.71 Fe II &  30 & 274 & 182 &.58  &   .11&-0.40&1.02\\
 5442.29*& 5442.26 Nd II &  75 & 100 &  63 &.61  &   .10&  &\\
         & 5442.37 Cr I  &     &     &     &         &        & &  \\
 5451.13*& 5451.12 Nd II &  93 & 207 &  65 &.85  &   .05&  &\\
         & 5451.13 Ce II &     &     &     &         &        & &  \\
 5454.24 & 5454.27 Er II &  65 & 497 & 196 &.14  &   .06&-1.4 &0.66\\
 5455.49*& 5455.44 Fe I  & 282 & 160 &  49 &.83  &   .05& &\\
         & 5455.47 Dy II &     &     &     &         &        & &  \\
 5456.55 & 5456.55 Nd II &  51 &  88 &  62 &.92  &   .11&-1.6 &1.25 \\
 5468.34 & 5468.37 Ce II &  50 & 145 &  41 &.92  &   .04&-1.9 &1.31\\
 5469.12 & 5469.10 Dy II &  53 & 133 &  45 &.79  &   .05&- &1.07\\
 5471.42 &               &  50 & 149 &  74 &.45  &   .08& &\\
 5482.30 & 5482.27 La II &  68 &  76 &  40 &.04  &   .08&-2.0 &1.45\\
 5487.65*& 5487.63 Mn I  & 188 &  32 &  25 &.95  &   .13&  &\\
         & 5487.74 Fe I  &     &     &     &         &        & &  \\
         & 5487.76 Fe I  &     &     &     &         &        &  & \\
 5513.44*& 5513.39 Fe I  &  80 & 136 &  46 &.00  &   .05& &\\
         & 5513.56 Pr II &     &     &     &         &        &  & \\
5525.109 & 5525.13 Fe II &  45 &  56 &  29 &.78  &   .08&-1.1 &0.1 \\
 5533.79*& 5533.79 Nd I  &  32 & 101 &  55 &.84  &   .09& &\\
         & 5533.82 V  I  &     &     &     &         &        & &  \\
 5543.15*& 5543.15 Fe I  & 100 & 110 &  35 &.75  &   .05& &\\
         & 5543.04 Cr II &     &     &     &         &        & &  \\

 5554.87*& 5554.89 Fe I  &  90 &  58 &  35 &.64  &   .10& &\\
& 5554.95 Cr I  &     &     &     &         &        & &  \\
 5557.90*& 5557.91 Fe I  &  80 & 133 &  55 &.91  &   .06&  &\\
         & 5557.98 Fe I  &     &     &     &         &        & &  \\
 5567.75*& 5567.75 Mn I  &  54 & 102 &  62 &.73  &   .10& &\\
         & 5567.84 Fe II &     &     &     &         &        & &  \\
 5569.07 & 5569.08 Cr II &  50 &  73 &  43 &.92  &   .09&-0.1 &1.07\\
\hline
\end{tabular}
\caption{Radial Velocity Amplitudes (cont.)}
\end{center}
\end{table}

\clearpage
\setcounter{table}{0}
\begin{table}
\begin{center}
\begin{tabular}{llccccccc}
Measured $\lambda$ & Identification &$W_{\lambda}$ & $K$ Amplitude & Error & Phase & Error&$log \tau$&$g_{eff}$ \\
{{\AA}} & {\hskip 20pt} {{\AA}} & m{{\AA}}& (m/s)  & (m/s)     &           &       & &\\
\hline
 5586.09*& 5586.05 Cr II &  55 & 163 &  58 &.89 &    .06& &\\
         & 5586.08 V I   &     &     &     &        &         & &  \\
         & 5586.13 Gd II &     &     &     &        &         & &  \\
 5586.75 & 5586.76 Fe I  & 130 &  76 &  30 &.70 &    .07& &\\
         & 5586.84 Cr I  &     &     &     &        &         & &  \\
 5588.92*& 5588.75 Ca I  & 140 &  44 &  21 &.64 &    .08&  &\\
         & 5588.93 Nd II &     &     &     &        &         & &  \\
 5594.55*& 5594.46 Ca I  & 188 &  70 &  46 &.84 &    .10& &\\
         & 5594.42 Nd II &     &     &     &        &         & &  \\
         & 5594.65 Fe I  &     &     &     &        &         &  & \\
 5602.84*& 5602.85 Ca I  & 230 &  16 &  30 &.94 &    .28& &\\
         & 5602.66 Nd II &     &     &     &        &         & &  \\
         & 5602.77 Fe I  &     &     &     &        &         &  & \\
         & 5602.88 Si I  &     &     &     &        &         &  & \\
         & 5602.94 Fe I  &     &     &     &        &         &  & \\
 5614.20 & 5614.28 Nd II &  47 & 155 &  94 &.89   &    .09&-1.9&0.94 \\
 5615.58* & 5615.64 Fe I & 132 &  60 &  19 &.76   &    .05&-2.7&1.19\\
         & 5615.30 Fe I  &     &     &     &        &         & &  \\
         & 5615.62 Fe I  &     &     &     &        &         &  & \\
 5620.54*& 5620.49 Fe I  & 121 & 109 &  40 &.74 &    .06 & &\\
         & 5620.59 Nd II &     &     &    &        &         & &  \\
         & 5620.63 Cr II &     &     &    &        &         &  & \\
 5625.63 & 5625.73 Nd II &  76 & 124 &  58 &.90   &    .07&-2.0&0.76\\
 5626.63*& 5626.73 Ce II &  77 & 147 &  43 &.01 &    .05& &\\
         & 5626.53 Er II &     &     &     &        &         & &  \\
         & 5626.73 Fe II &     &     &     &        &         &  & \\
 5630.35 & 5630.38 Ce II &  38 & 192 &  57 &.02 &    .05&  &\\
 5637.47*& 5637.36 Ce II &  85 &  77 &  41 &.79 &    .09& &\\
         & 5637.30 Sm II &     &     &     &      &           && \\
 5638.33*& 5638.27 Fe I &  67 & 139 &  55 &.88 &    .06 &-1.1&1.12 \\
         & 5638.16 Cr I  &     &     &     &      &           & &\\
 5640.45 & uncl          &  30 &  75 &  58 &.42 &    .12& &\\
 5641.61*& 5641.69 Cr I  & 109&   70 &  48 &.94 &    .11& &\\
         & 5641.50 Dy II &    &      &     &      &           & & \\
 5650.32 & uncl          &  48 &  69 &  32 &.89 &    .07& &\\
 5658.42*& 5658.53 Fe I  & 150 &   4 &  21 &.61 &    .84& &\\
         & 5658.36 Sc II &     &     &     &      &           & &\\
 5678.46*& 5678.39 Cr II & 167 & 286 & 113 &.49 &    .06& &\\
         & 5678.61 Cr I  &     &     &     &      &           & &\\
 5679.56 & 5679.59 Sm II &  70 & 160 &  59 &.05 &    .06& &\\
 5680.22*& 5680.26 Ce II &  70 &  22 &  33 &.73 &    .24& &\\
         & 5680.24 Fe I  &     &     &     &      &           && \\
\hline
\end{tabular}
\caption{Radial Velocity Amplitudes Lines (cont.)}
\end{center}
\end{table}

\setcounter{table}{0}
\clearpage
\begin{table}
\begin{center}
\begin{tabular}{llccccccc}
Measured $\lambda$ & Identification &$W_{\lambda}$ & $K$ Amp. & Error  & Phase  & Error & $log \tau$&$g_{eff}$\\
{{\AA}} & {\hskip 20pt} {{\AA}} & m{{\AA}}& (m/s)  & (m/s)     &        &        &&\\
\hline
 5691.48*& 5691.46 Ce II &  32 & 188 &  67 &.16&  .06 & &\\
         & 5691.48 Ni I  &     &     &     &     &          && \\
         & 5691.50 Fe I  &     &     &     &     &          && \\
 5721.96 & 5721.96 Gd II &  72 & 174 &  53 &.04&  .05 &-1.8&1.06 \\
 5753.09*& 5753.12 Fe I  &  84 & 100 &  45 &.76&  .07 &-1.2 &0.94\\
         & 5753.02 Pr II &     &     &     &     &          & &\\
 5757.65 & 5757.63 Er II &  61 & 214 &  66 &.98&  .05 & &\\
 5761.62 & 5761.69 Nd II &  49 & 114 &  45 &.83 &  .06&-1.9&0.91\\
 5763.00*& 5762.99 Fe I  & 122 &  70 &  39 &.88&  .09 & &\\
         & 5762.99 Ce II &     &     &     &     &          & &\\
         & 5762.98 Si I  &     &     &     &     &          &  &\\
 5804.07*& 5804.00 Nd II & 103 & 175 &  48 &.81 &  .04&  &\\
         & 5804.03 Fe I  &     &     &     &      &         &  &\\
 5834.07& 5833.93 Fe I  &  70 & 203 &  57 &.02 &  .04& -1.0 &1.99\\
 5835.45*& 5835.49 Fe II &  35 & 165 &  53 &.63 &  .06&& \\
         & 5835.42 Fe I  &     &     &     &      &         &  &\\
 5859.67*& 5859.58 Fe I  &  93 &  52 &  27 &.91 &  .08&  &\\
         & 5859.67 Pr II &     &     &     &      &         & & \\
 5998.90 & 5998.94 Pr III& 109 & 105 &  64 &.21 &  .10&   &\\
 6004.62 & 6004.56 Gd II &  70 & 106 &  41 &.95 &  .06&-2.4&1.06 \\
 6014.54 & uncl          &  76 & 125 &  61 &.50 &  .07&  &\\
 6024.05 & 6024.06 Fe I  &  90 &  76 &  60 &.83 &  .13&-1.6 &1.17\\
 6027.03*& 6027.05 Fe I  &  54 &  98 &  60 &.84 &  .10&-1.1&0.96\\
         & 6027.05 Cr II &     &     &     &      &         &  &\\

 6034.18*& 6034.23 Nd II &  64 & 248 &  50 &.80 &  .03&-2.1  &1.24\\
         & 6034.21 Ce II &     &     &     &      &         & & \\
 6051.85*& 6051.86 Nd II &  54 & 318 &  69 &.06 &  .03&-1.7  &1.23\\
         & 6051.80 Ce II &     &     &     &      &         & & \\
 6065.49 & 6065.48 Fe I  &  65 &  50 & 43  &.92 &  .13&-1.9&0.68\\
 6122.29 & 6122.22 Ca I  & 176 &  66 &  37 &.65 &  .09&-3.4&1.75\\
 6127.90 & 6127.91 Fe I  &  30 &  93 &  71 &.56 &  .12& -0.9  &0.72\\
 6136.62*& 6136.61 Ti I  &  90 & 104 &  87 &.45 &  .13&      &\\
         & 6136.62 Fe I  &     &     &     &     &       &      &\\
         & 6136.63 Cr I  &     &     &     &      &       &      &\\
 6140.30 & 6140.22 Ce II &  32 & 130 &  90 &.07 &  .11&   &\\
\hline
\end{tabular}
\caption{Radial Velocity Amplitudes (cont.)}
\end{center}
\end{table}

\clearpage
\setcounter{table}{0}
\begin{table}
\begin{center}
\begin{tabular}{llccccccc}
Measured $\lambda$ & Identification &$W_{\lambda}$ & $K_o$ Amplitude & Error & Phase & Error& $log \tau$ &$g_{eff}$\\
{{\AA}} & {\hskip 20pt} {{\AA}} & m{{\AA}}& (m/s)  & (m/s)     &  &  &&\\
\hline
 6141.71*& 6141.71 Ba II & 130 & 198 &  92 &.11&    .07 &&\\
         & 6141.73 Fe I  &     &     &     &     &            &&\\
 6145.05*& 6145.07 Nd III& 190 & 144 &  36 &.42&    .04 &-2.4&\\
         & 6145.02 Si I  &     &     &     &     &            &&\\
 6147.75*& 6147.74 Fe II & 115 & 168 &  85 &.81 &    .08& &\\
         & 6147.83 Fe I  &     &     &     &     &            &&\\

 6230.75*& 6230.72 Fe I  &  80 & 440 & 120 &.83 &    .04& &\\
         & 6230.80 V I   &     &     &     &     &            &&\\
\end{tabular}
\caption{Radial Velocity Amplitudes (cont).}
\end{center}
\end{table}
\clearpage

\end{document}